\documentclass[12pt]{article}
\usepackage{graphicx}
\usepackage{epstopdf}
\usepackage{amssymb,amsmath,epsfig}

\begin{document}

\title{\bf Cracking of Charged Polytropes with Generalized Polytropic Equation of State}

\author{M. Azam$^{1}$
\thanks{azam.math@ue.edu.pk} and S. A. Mardan$^2$ \thanks{syedalimardanazmi@yahoo.com}\\
$^1$ Division of Science and Technology, University of Education,\\
Township Campus, Lahore-54590, Pakistan.\\
$^2$ Department of Mathematics,\\ University of the Management and Technology,\\
C-II, Johar Town, Lahore-54590, Pakistan.}

\date{}

\maketitle
\begin{abstract}
We discuss the occurrence of cracking in charged anisotropic polytropes
with generalized polytropic equation of state
through two different assumptions; (i) by carrying out local density perturbations under
conformally flat condition (ii) by perturbing anisotropy, polytropic index
and charge parameters. For this purpose, we consider two different definitions
of polytropes exist in literature. We conclude that under local density
perturbations scheme cracking does not appears in both types of polytropes
and stable configuration are observed, while with second kind of perturbation
cracking appears in both types of polytropes under certain conditions.
\end{abstract}
\section{Introduction}

The theory of polytropes is vital in the evolution of
mathematical models of compact objects and it attracts many researchers
due to its simple form. In the study of polytropes,
the main attraction is Lane-Emden equation, which led us
towards the illustration of various astrophysical phenomena. Chandrasekhar \cite{1}
initially developed the theory of polytropes in Newtonian frame work
with the help of laws of thermodynamics. Topper \cite{2,3} used the hypothesis
of quasi-static equilibrium form for the development of initial frame work
of Newtonian polytropes. Kovetz \cite{4} refined the work of Chandrasekhar \cite{1}
and reshaped the theory of polytropes. Abramowicz \cite{5} was the first who
presented higher dimension polytropes by developing Lane-Emden equation in higher dimension.

The study of electromagnetism and its effect on physical properties
of astrophysical objects always fascinated the researches. Bekenstein \cite{6}
developed hydrostatic equilibrium equation (HEe) for the description of collapse
of charged stars. Bonnor \cite{7,8} presented the study on charged compact objects
and described how electromagnetism affects the gravitational collapse.
Bondi \cite{9} used isotropic coordinates to analyzed the contraction
of stars in the presence of charged. 
Ray et al. \cite{11} examined the properties of stars with higher densities and
concluded that approximately $10^{20}$ coulomb charge can be hold by such stars.
Herrera et al. \cite{12} utilized structure scalars to illustrate compact objects having
charged dissipative inner fluid distribution. Takisa and Maharaj
\cite{13} presented the mathematical model of charged compact objects with polytropic EoS.

The impact of anisotropy in the theory of general relativity is very important as
we cannot study many physical phenomena without taking it into account.
Cosenza et al. \cite{14} presented a heuristic way for mathematical modeling
of compact objects with anisotropic inner fluid distribution.
Herrera and Santos \cite{15} derived the anisotropic compact models
in the frame work of general relativity.
Herrera and Barreto \cite{16} adopted a novel approach of effective variables
for the description of physical variables involved in the anisotropic
polytropic models. Herrera et al. \cite{17} developed the governing equations in the presence of
anisotropic stress for spherically symmetries.
Herrera and Barreto \cite{18,19} used the concept of Tolman-mass to check the
viability of anisotropic polytropic models. Herrera et al. \cite{20} adopted
conformally flat approach to reduced physical parameters for
simplification of Lane-Emden equations of polytropes.

The stability analysis of any developed model of stars plays a
key role in general relativity. Any developed model cannot be
used for the description of stars unless it is critically
analyzed for stability. Bondi \cite{21} developed HEe for stability
analysis of neutral stars. Herrera et al. \cite{22} proposed a novel way for
the analysis of spherical symmetric models by means of cracking
(overturning), which described the behavior of fluid distribution
just after equilibrium state has been
perturbed through density perturbation. Gonzalez et al. \cite{23,24}
 provided an extension of Herrera et al. \cite{22} by introducing
 local density perturbation (LDP). Azam et al. \cite{25}-\cite{29} used
 LDP for the analysis of various mathematical models of compact objects.
 Sharif and Sadiq \cite{30} developed the model of charged
polytropes. Azam et al. \cite{31,32} developed the general frame work
for charged polytropes with generalized polytropic equation of state
(GPEoS) for spherical and cylindrical symmetries. They analyzed these
models by means of Tolman-mass and Whittaker mass for spherical and cylindrical symmetries, respectively.
Herrera et al. \cite{33} have discussed the effect of small fluctuations of local anisotropy of pressure, and energy density
on spherical polytropes. Sharif and Sadiq \cite{34} have examined the effects of charge
on spherical polytropes. Azam and Mardan \cite{35}
refined the work \cite{34} for the analysis of charged polytropes.

The plan of work is as follows. In section \textbf{2}, we provide some basic equations. Section \textbf{3}
and \textbf{4} are devoted for the analysis of cracking through LDP and parametric perturbation respectively.
In the last section we conclude over results.

\section{Einstein-Maxwell Field Equations}
{
\allowdisplaybreaks
We consider static spherically symmetric space time
\begin{equation}\label{1}
ds^2=-e^{\nu}dt^{2}+e^{\lambda}dr^{2}+r^2 d\theta^{2}+r^2 \sin^2\theta{d\phi^2},
\end{equation}
where $\nu(r)$ and $\lambda(r)$ both depends only on radial
coordinate $r$. The generalized form of energy-momentum tensor
for charged anisotropic inner fluid distribution is given by
\begin{equation}\label{2}
T_{i j}=(P_t+\rho) V_{i} V_{j} +g_{i j}P_t +(P_r - P_t) S_{i}
S_{j}+\frac{1}{4\pi}(F_{i}^{m} F_{j m}-\frac{1}{4} F^{mn} F_{mn} g_{ij}),
\end{equation}
where $P_t$, $P_r$, $\rho$, $V_i$, $S_i$ and $F_{mn}$ represent the
tangential pressure, radial pressure, energy density, four velocity,
four vector and Maxwell field tensor for the inner fluid distribution.
The Einstein-Maxwell field equations for line element Eq.$(\ref{1})$
are given by
\begin{eqnarray}\label{7}
\frac{\lambda^\prime
e^{-\lambda}}{r}+\frac{(1-e^{-\lambda})}{r^2}=8\pi \rho
+\frac{q^2}{r^4}, \\\label{8}
\frac{\nu^\prime e^{-\lambda}}{r}-\frac{(1-e^{-\lambda})}{r^2}=8\pi P_r
-\frac{q^2}{r^4}, \\\label{9}
e^{-\lambda} \bigg[\frac{\nu^{\prime\prime}}{2}-\frac{\nu^\prime
\lambda^\prime}{4}+\frac{\nu^{\prime^2}}{4}+\frac{\lambda^\prime
-\nu^\prime}{2r}\bigg] = 8 \pi P_t+\frac{q^2}{r^4}.
\end{eqnarray}
Solving Eqs. $(\ref{7})$-$(\ref{9})$ simultaneously lead to HEe
\begin{equation}\label{10}
\frac{d P_r}{dr}-\frac{2}{r}\Big(\Delta+\frac{q q^\prime}
{8\pi r^3}\Big)+\Big(\frac{4\pi r^4 P_r-q^2+m r}{r(r^2-2mr+q^2)}\Big)(\rho+P_r)=0,
\end{equation}
where we have used $\Delta=(P_t-P_r)$.
We take the Reissner-Nordstr\"{a}m space-time as the exterior geometry
\begin{equation}\label{11}
ds^2=\Big(1-\frac{2M}{r}+\frac{Q^2}{r^2}\Big)dt^2-
\Big(1-\frac{2M}{r}+\frac{Q^2}{r^2}\Big)^{-1}dr^2-r^2
d\theta^{2}-r^2 \sin^2\theta{d\phi^2}.
\end{equation}
The junction conditions are very important in mathematical modeling of
compact stars. They provide us the criterion for the
collaboration of two metrics, which can results a physically viable
solution \cite{35,36}. For smooth matching of two space times, we must have
\begin{equation}\label{12}
e^{\nu}=e^{-\lambda}=\Big(1-\frac{2M}{r}+
\frac{Q^2}{r^2}\Big),~~m(r)=M,~q(r)=Q,~P_r=0,
\end{equation}
and Misner-Sharp mass \cite{37} leads to
\begin{equation}\label{13}
m(r)=\frac{r}{2}(1-e^{-\lambda}+\frac{q^2}{r^2}),
\end{equation}
which has been used in the development of Lane-Emden equations \cite{31}.

\section{Effect of Local Density Perturbation}

In this section, we apply LDP \cite{23,24} on charged
conformally flat polytropes in equilibrium state.
In LDP scheme it is assumed that all the physical
parameters involved in the model and their derivative
as function of density. Then the density is perturbed
slightly and its effects has been observed on the HEe.
Two different kinds of polytropes exist in literature will discussed here.

\subsection{Case 1}
We consider the GPEoS as
\begin{equation}\label{14}
P_r=\alpha_1\rho_o+K\rho_o^{\gamma}=\alpha_1\rho_o+K\rho_o^{1+\frac{1}{n}},
\end{equation}
so that the original polytropic part remain conserved,
also the mass density $\rho_o$ is related to total energy density
$\rho$ as \cite{7}
\begin{equation}\label{15}
\rho=\rho_{o}+n P_r.
\end{equation}
Now taking following assumptions
\begin{equation*}
\alpha=\frac{P_{rc}}{\rho_{gc}},~~~\alpha_2
=1+(n+1)(\alpha_1+\alpha\theta),~~~\alpha_3
=(n+1)\alpha,~~~\alpha_4=\frac{4\pi P_{rc} q^2}{\alpha \alpha_3},
\end{equation*}
\begin{equation}\label{16}
r=\frac{\xi}{A},~~~\rho_{o}=\rho_{gc}\theta^n,~~~m(r)
=\frac{4\pi\rho_{gc} v({\xi})}{A^3},~~~A^2=
\frac{4\pi\rho_{gc}}{(n+1)\alpha},
\end{equation}
where $P_{rc}$ is the pressure at center of the star,
$\rho_{gc}$ is the mass density
evaluated at the center of CO, $\xi$, $\theta$ and
$v$ are dimensionless variables.
We use conformally flat condition to
find the expression of anisotropy factor $\Delta$.
The electric part of Weyl tensor is related to Weyl scalar given by \cite{7,17}
\begin{equation}\label{17}
W=\frac{r^3 e^{-\lambda}}{6}\Bigg(\frac{e^\lambda}{r^2}
+\frac{\lambda^\prime \nu^\prime}{4}
-\frac{1}{r^2} -\frac{\nu^{\prime 2}}{4}-
\frac{\nu^{\prime\prime}}{2}-\frac{\lambda^\prime \nu^\prime}{2 r}\Bigg).
\end{equation}
Using conformally flat condition, i.e., $W=0$, along with Eqs.$(\ref{7})$-$(\ref{9})$ in Eq.$(\ref{17})$, we get
\begin{equation}\label{18}
\Delta=P_t-P_r=\frac{e^{-\lambda}}{4\pi}\Bigg(\frac{e^{\lambda}}{r^2}
-\frac{\lambda^\prime}{2r} -\frac{1}{r^2}\Bigg)-\frac{q^2}{4 \pi r^4}.
\end{equation}
Now differentiating Eq.$(\ref{13})$ with respect to $``r"$ and using the
assumptions given in Eq.$(\ref{16})$, we get
\begin{eqnarray}\label{19}
\frac{dv(\xi)}{d\xi}=\xi^2\theta^n(1+n\alpha_1+n\alpha\theta)
-\frac{\alpha_4}{\alpha_3\xi^2}+\frac{\alpha_4}
{\alpha_3 \xi q}\frac{d q}{d\xi}.
\end{eqnarray}
The above equation along with Eq.$(\ref{16})$ and $(\ref{18})$ yields
\begin{equation}\label{20}
\Delta=\rho_{gc}\Bigg((1+n\alpha_1+n\alpha\theta)
\theta^n+3\frac{v(\xi)}{\xi^3}
-4\frac{\alpha_4}{\alpha_3 \xi^4}+2
\frac{\alpha_4 \frac{d q}{d \xi}}{\alpha_3 q \xi^3}\Bigg).
\end{equation}
In order to observe the effects of LDP on conformally flat polytropes,
we transform the HEe Eq.$(\ref{10})$
by using Eqs.$(\ref{14})$, $(\ref{16})$ and $(\ref{20})$ as
\begin{eqnarray}\label{21}\notag
&&R_1 \approx(n\alpha_1 \theta^{n-1}+\alpha_3\theta^{n})\frac{d \theta}{d \xi}
-\frac{2}{\xi}\Big\{(\alpha_2-\alpha_1-\alpha\theta)
\theta^{n} +3\frac{v(\xi)}{\xi^3}
-4\frac{\alpha_4}{\alpha_3 \xi^4}\\ \notag
 &&+2 \frac{\alpha_4 }{\alpha_3 q \xi^3}\frac{d q}{d \xi} \Big\}
 +(\alpha_2-1)\Big\{\frac{(\alpha_1+\alpha \theta)\theta^{n}\xi^4
 - \alpha_3^{-1}\alpha_4+v(\xi)\xi}{\xi^3\alpha_3^{-1}
 -2v(\xi)\xi+\alpha_3^{-1}\alpha_4 \xi }\Big\}\theta^{n}\\
 &&-\frac{\alpha_4 }{\alpha_3 q \xi^4}\frac{d q}{d \xi}=0.
\end{eqnarray}
Now we apply LDP to perturb all the physical variables in Eq.$(\ref{21})$
and for this purpose we can write
\begin{eqnarray}\label{22}
\theta(\rho_{gc}+\delta\rho_{gc})&=&\theta(\rho_{gc})
+\frac{d \theta}{d\rho_{gc}}\delta\rho_{gc}=\theta(\rho_{gc})+
 \frac{\frac{d \theta}{d\xi}}{\frac{d \rho_{gc}}{d\xi}}\delta\rho_{gc},
\\
\label{23}
\frac {d \theta }{d \xi}(\rho_{gc}+\delta\rho_{gc})&=&\frac {d \theta }{d \xi}(\rho_{gc})
+   \frac{\frac {d^2 \theta }{d \xi^2}}{\frac{d \rho_{gc}}{d\xi}} \delta\rho_{gc},
\\
 \label{24}
 v(\rho_{gc}+\delta\rho_{gc})&=&
 v(\rho_{gc})+
 \frac{\frac {d v}{d \xi}}{\frac{d \rho_{gc}}{d\xi}} \delta\rho_{gc},
 \\
 \label{25}
 P_{rc}(\rho_{gc}+\delta\rho_{gc})&=&
 P_{rc}(\rho_{gc})+
 \frac{\frac {d P_{rc}}{d \xi}}{\frac{d \rho_{gc}}{d\xi}} \delta\rho_{gc},
\\
\label{26} q(\rho_{gc}+\delta\rho_{gc})&=&
q(\rho_{gc})+
 \frac{\frac {d q}{d \xi}}{\frac{d \rho_{gc}}{d\xi}} \delta\rho_{gc},
 \\
\label{27}
\frac {d q }{d \xi}(\rho_{gc}+\delta\rho_{gc})&=&\frac {d q }{d \xi}(\rho_{gc})
+   \frac{\frac {d^2 q }{d \xi^2}}{\frac{d \rho_{gc}}{d\xi}} \delta\rho_{gc}.
\end{eqnarray}
So the perturb form of the Eq.$(\ref{21})$ can be written as
\begin{eqnarray}\label{28}
R_1=R_1 (\theta,\frac {d \theta }{d \xi}, v, P_{rc}, q, \frac {d q }{d \xi}, \rho_{gc} )+\delta R_1,
\end{eqnarray}
where
\begin{eqnarray}\label{29} \notag
\delta R_1 &=& \Big(\frac{d \rho_{gc}}{d \xi}\Big)^{-1}
\Bigg\{\frac{\partial R_1}{\partial \rho_{gc}}\delta\rho_{gc} +
            \frac{\partial R_1}{\partial \theta} \frac{d\theta}{d \xi}   +
            \frac{\partial R_1}{\partial \frac {d \theta }{d \xi}} \frac{d^2\theta}{d \xi^2}     +
            \frac{\partial R_1}{\partial v}\frac{d v}{d \xi}  \\&+&
            \frac{\partial R_1}{\partial P_{rc}}\frac{d P_{rc}}{d \xi}   +
            \frac{\partial R_1}{\partial q} \frac{d q}{d \xi}   +
            \frac{\partial R_1}{\partial \frac {d q }{d \xi}}
            \frac{d^2 q }{d \xi^2}  \Bigg\}\delta \rho_{gc}.
\end{eqnarray}
We will plot the force distribution
$\frac{\delta R_1}{\delta \rho_{gc}}$ against the
dimensionless radius $\xi$ to observe possible occurrence of cracking. We say that
cracking appears if force distribution changes it sign.

\subsection{Case 2}
Here, we consider the GPEoS as
\begin{equation}\label{30}
P_r=\alpha_1\rho+K\rho^{1+\frac{1}{n}},
\end{equation}
where mass density $\rho_o$ is replaced by
total energy density $\rho$ in Eq.$(\ref{14})$ and
they are related to each other as \cite{7}
\begin{equation}\label{31}
\rho=\frac{\rho_o}{\big(1-K \rho_o^{\frac{1}{n}}\big)^n}.
\end{equation}
We take following assumptions
\begin{equation*}
\alpha=\frac{P_{rc}}{\rho_{c}},~~~\alpha_5=1+\alpha_1+\alpha\theta,
\end{equation*}
\begin{equation}\label{32}
r=\frac{\xi}{A},~~~\rho_{o}=\rho_{c}\theta^n,~~~m(r)
=\frac{4\pi\rho_{c}
v({\xi})}{A^3},~~~A^2=\frac{4\pi\rho_{c}}{(n+1)\alpha},
\end{equation}
where $c$ represents the quantity at center of the star,
$\alpha_2$, $\alpha_3$ and $\alpha_4$
 are same expressions as in Eq.$(\ref{16})$
with $\alpha$ defined in Eq.$(\ref{32})$.
Carrying out the same process as in case 1, we get
\begin{eqnarray}\label{33}
\frac{dv(\xi)}{d\xi}=\xi^2\theta^n-\frac{\alpha_4}{\alpha_3
\xi^2}+\frac{\alpha_4}{\alpha_3 \xi q}\frac{d q}{d\xi},
\end{eqnarray}
and the anisotropy factor $\Delta$ comes out to be
\begin{equation}\label{34}
\Delta=\rho_{c}\Bigg(\theta^n+3\frac{v(\xi)}{\xi^3}
-4\frac{\alpha_4}{\alpha_3 \xi^4}+2\frac{\alpha_4 \frac{d q}{d
\xi}}{\alpha_3 q \xi^3}\Bigg).
\end{equation}
The HEe Eq.$(\ref{10})$ will transform as
\begin{eqnarray}\label{35}\notag
&&R_2 \approx(n\alpha_1 \theta^{n-1}+\alpha_3\theta^{n})\frac{d \theta}{d \xi}
-\frac{2}{\xi}\Big\{(1+\alpha_2-\alpha_5+n(\theta_1+\alpha\theta))\theta^{n} \\ \notag
 &&+3\frac{v(\xi)}{\xi^3}
-4\frac{\alpha_4}{\alpha_3 \xi^4}+2
\frac{\alpha_4 }{\alpha_3 q \xi^3}\frac{d q}{d \xi} \Big\}
 -\frac{\alpha_4 }{\alpha_3 q \xi^4}\frac{d q}{d \xi}
 \\
 &&
 +\alpha_5\Big\{\frac{(\alpha_1+\alpha \theta)
 \theta^{n}\xi^4 - \alpha_3^{-1}\alpha_4+v(\xi)\xi}
 {\xi^3\alpha_3^{-1}-2v(\xi)\xi+\alpha_3^{-1}\alpha_4 \xi }\Big\}\theta^{n}=0,
\end{eqnarray}
proceeding in the same way, the perturb form of the Eq.$(\ref{35})$ can be written as
\begin{eqnarray}\label{36}
R_2=R_2 (\theta,\frac {d \theta }{d \xi}, v, P_{rc},
 q, \frac {d q }{d \xi}, \rho_{gc} )+\delta R_2,
\end{eqnarray}
where
\begin{eqnarray}\label{37} \notag
\delta R_2 &=& \Big(\frac{d \rho_{gc}}{d \xi}\Big)^{-1}
\Bigg\{\frac{\partial R_2}{\partial \rho_{gc}}\delta\rho_{gc} +
            \frac{\partial R_2}{\partial \theta} \frac{d\theta}{d \xi}   +
            \frac{\partial R_2}{\partial \frac {d \theta }{d \xi}} \frac{d^2\theta}{d \xi^2}     +
            \frac{\partial R_2}{\partial v}\frac{d v}{d \xi}  \\&+&
            \frac{\partial R_2}{\partial P_{rc}}\frac{d P_{rc}}{d \xi}   +
            \frac{\partial R_2}{\partial q} \frac{d q}{d \xi}   +
            \frac{\partial R_2}{\partial \frac {d q }{d \xi}}\frac{d^2 q }{d \xi^2}  \Bigg\}\delta \rho_{gc}.
\end{eqnarray}
We will plot the force distribution
$\frac{\delta R_2}{\delta \rho_{gc}}$ against the
dimensionless radius $\xi$ to observe possible
occurrence of cracking. We say that cracking
appears if force distribution changes it sign.
\begin{figure} \label{fig1}
\centering
\includegraphics[width=80mm]{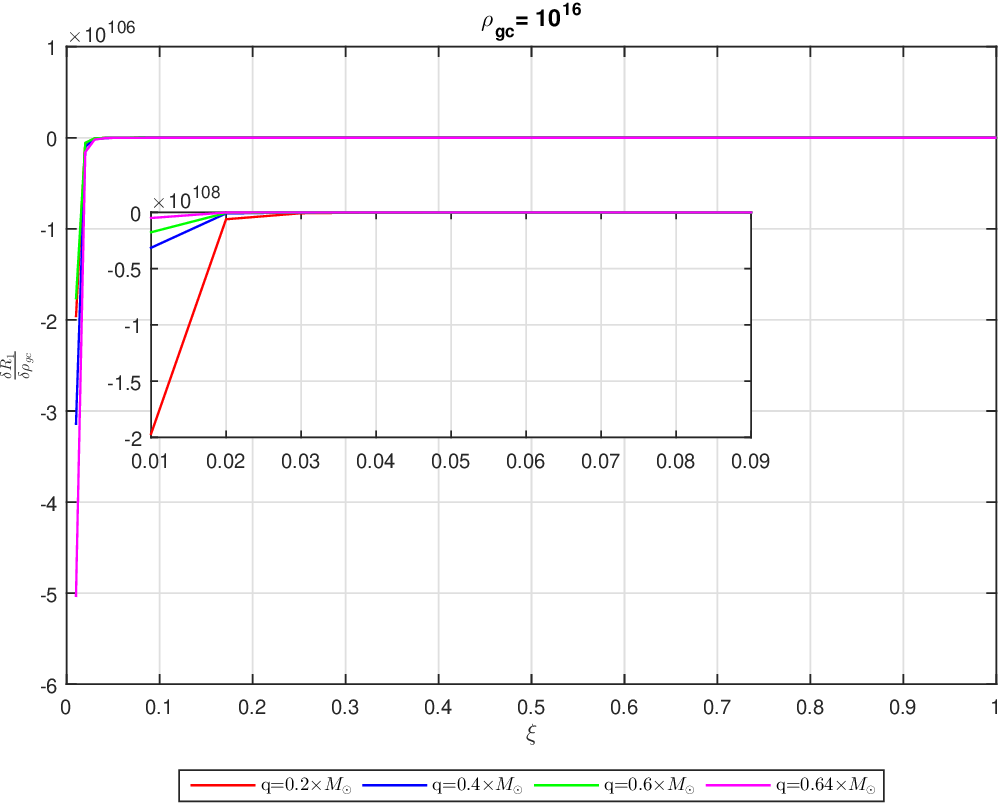}
\caption{Case $1$: Perturbation through LDP.
$\frac{\delta R_1}{\delta \rho_{gc}}$ as a function of $\xi$ for $n=1,~\alpha=8 \times 10 ^{-11},~\alpha_1=0.2$,
red curve: q=0.2 $M_\odot$,
blue curve: q=0.4 $M_\odot$,
green curve: q=0.6 $M_\odot$,
magenta curve: q=0.64 $M_\odot$}.
\end{figure}
\begin{figure} \label{fig2}
\centering
\includegraphics[width=80mm]{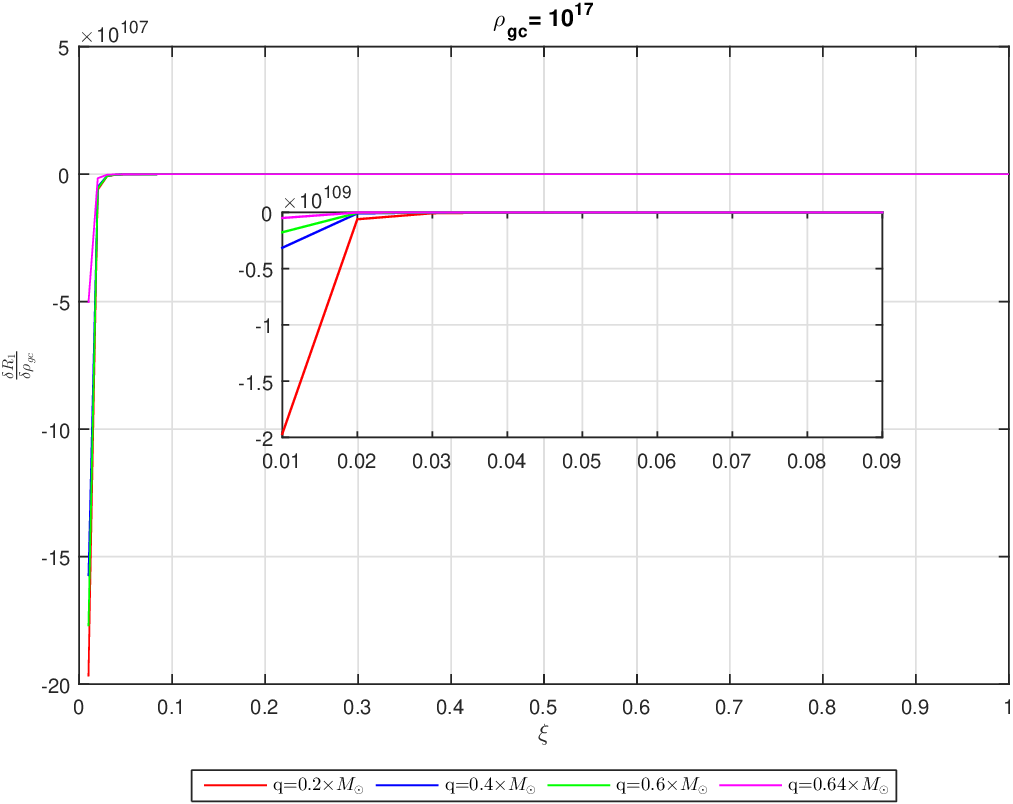}
\caption{Case $1$: Perturbation through LDP. $\frac{\delta R_1}{\delta \rho_{gc}}$ as a function of $\xi$ for $n=1.5,~\alpha=2 \times 10 ^{-10},~\alpha_1=0.5$,
red curve: q=0.2 $M_\odot$,
blue curve: q=0.4 $M_\odot$,
green curve: q=0.6 $M_\odot$,
magenta curve: q=0.64 $M_\odot$}.
\end{figure}
\begin{figure} \label{fig3}
\centering
\includegraphics[width=80mm]{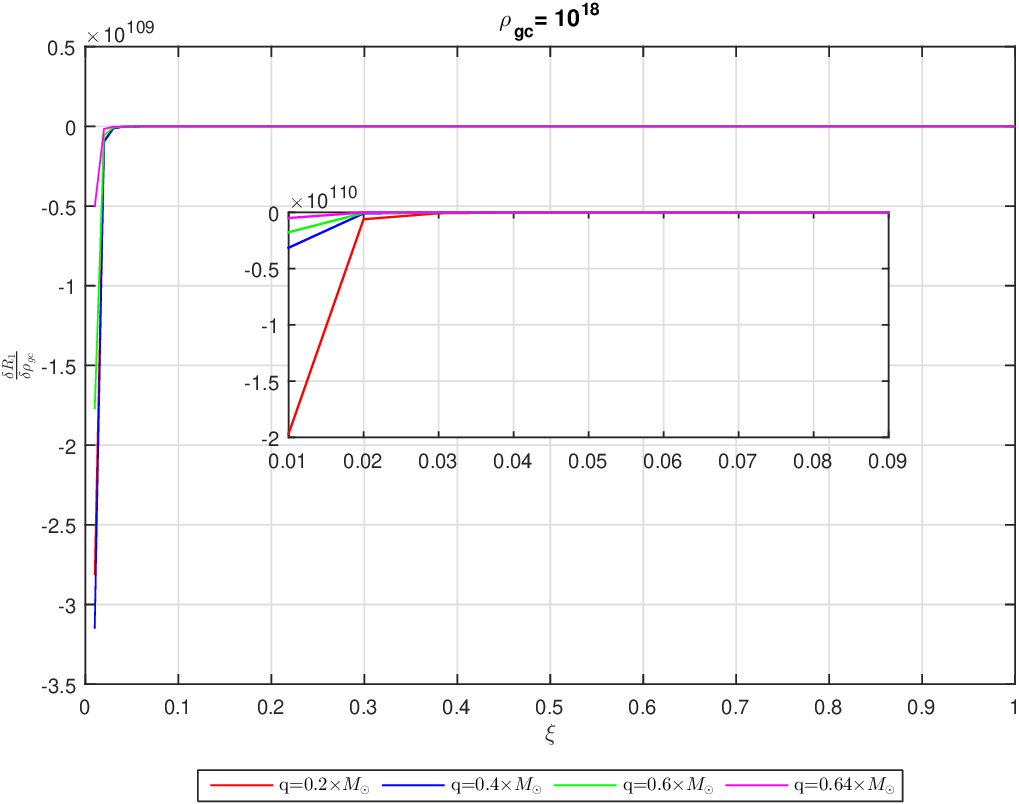}
\caption{Case $1$: Perturbation through LDP. $\frac{\delta R_1}{\delta \rho_{gc}}$ as a function of $\xi$ for $n=1,~\alpha=0.5,~\alpha_1=0.8$,
red curve: q=0.2 $M_\odot$,
blue curve: q=0.4 $M_\odot$,
green curve: q=0.6 $M_\odot$,
magenta curve: q=0.64 $M_\odot$}.
\end{figure}
\begin{figure} \label{fig4}
\centering
\includegraphics[width=80mm]{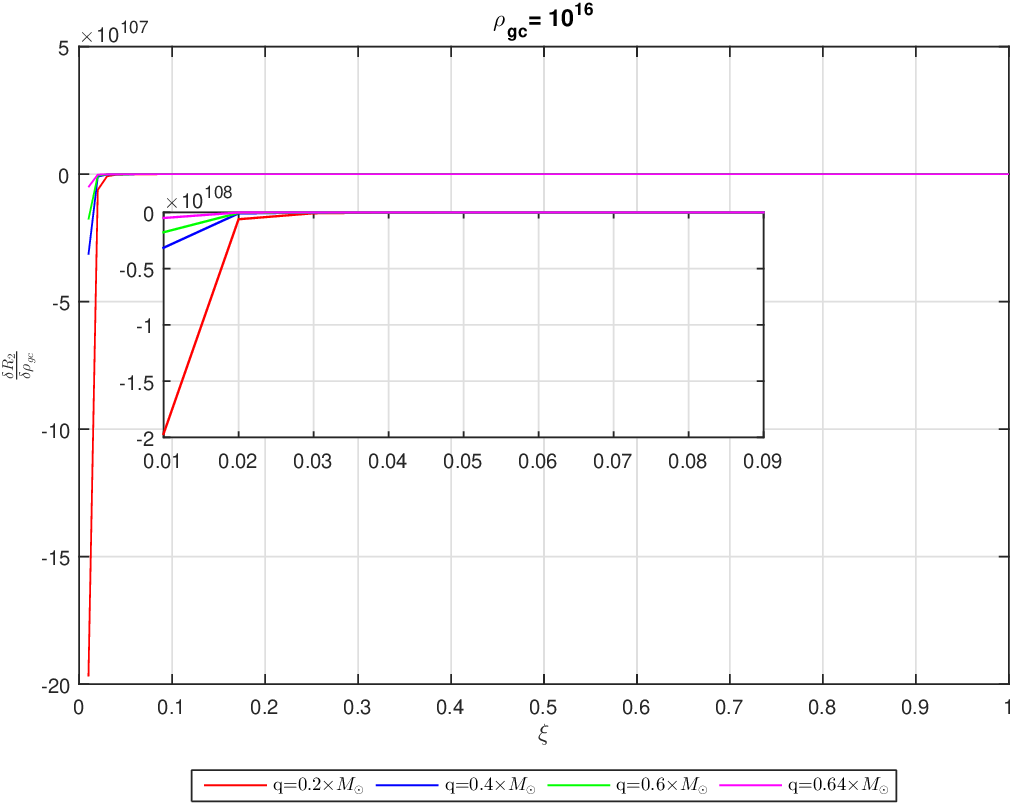}
\caption{Case $2$: Perturbation through LDP. $\frac{\delta R_2}{\delta \rho_{gc}}$ as a function of $\xi$ for $n=1,~\alpha=8 \times 10 ^{-11},~\alpha_1=0.2$,
red curve: q=0.2 $M_\odot$,
blue curve: q=0.4 $M_\odot$,
green curve: q=0.6 $M_\odot$,
magenta curve: q=0.64 $M_\odot$}.
\end{figure}
\begin{figure} \label{fig5}
\centering
\includegraphics[width=80mm]{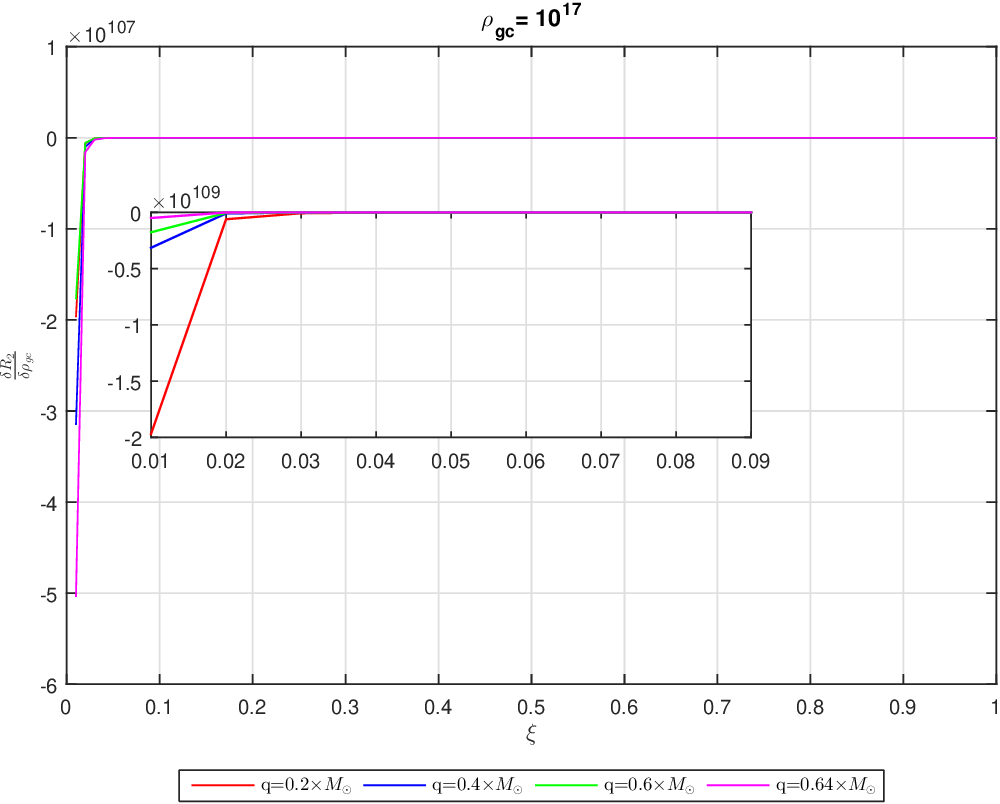}
\caption{Case $2$: Perturbation through LDP. $\frac{\delta R_2}{\delta \rho_{gc}}$ as a function of $\xi$ for $n=1.5,~\alpha=2 \times 10 ^{-10},~\alpha_1=0.5$,
red curve: q=0.2 $M_\odot$,
blue curve: q=0.4 $M_\odot$,
green curve: q=0.6 $M_\odot$,
magenta curve: q=0.64 $M_\odot$}.
\end{figure}
\begin{figure} \label{fig6}
\centering
\includegraphics[width=80mm]{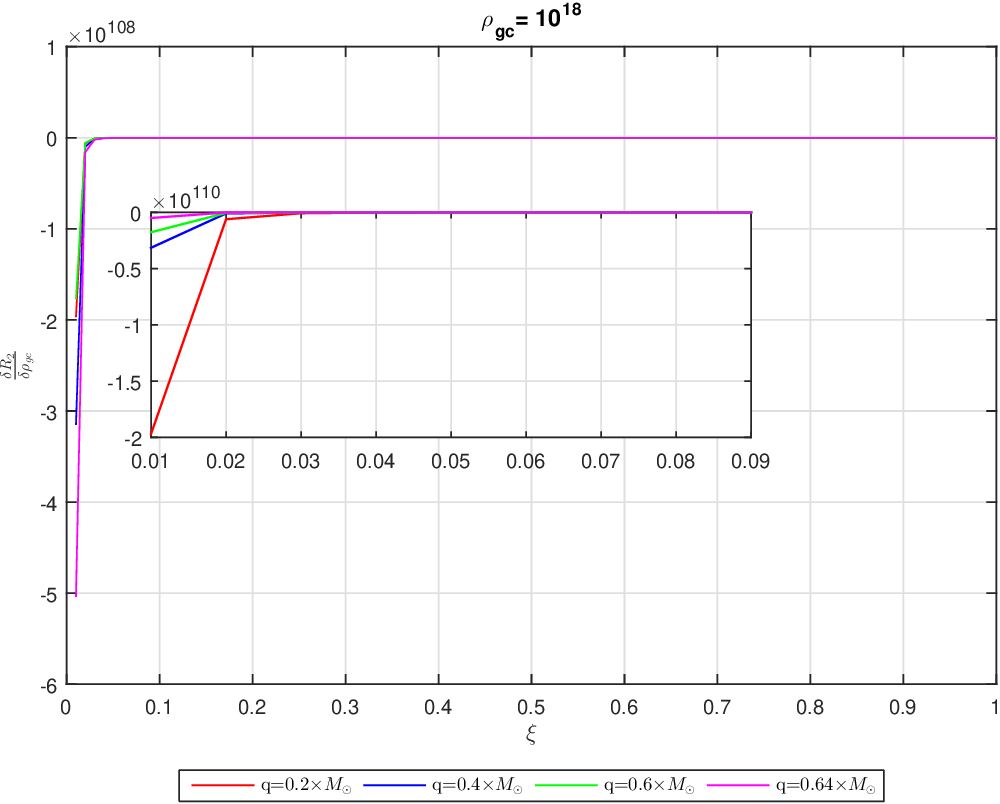}
\caption{Case $2$: Perturbation through LDP. $\frac{\delta R_2}{\delta \rho_{gc}}$ as a function of $\xi$ for $n=1,~\alpha=0.5,~\alpha_1=0.8$,
red curve: q=0.2 $M_\odot$,
blue curve: q=0.4 $M_\odot$,
green curve: q=0.6 $M_\odot$,
magenta curve: q=0.64 $M_\odot$}.
\end{figure}

\section{Effect of Parametric Perturbation}

In this section, we will study the stability of charged polytropes
by perturbing the polytropic index and anisotropy factor. For this
purpose we assume that our distribution satisfy the following relation
\begin{equation}\label{38}
\Delta=C(\rho+P_r)\Big[\frac{4\pi r^4 P_r-q^2+m r}
{r^2-2mr+q^2}\Big],
\end{equation}
where $C$ is constant, producing the following
form of HEe
\begin{equation}\label{39}
R_3=\frac{d P_r}{dr}+h(\rho+P_r)\Big(\frac{4\pi r^4 P_r-q^2+m r}
{r(r^2-2mr+q^2)}\Big)-\frac{q}{4\pi r^4}\frac{d q}{dr},
\end{equation}
with $h=1-2C$.
Now we shall briefly review the main results for each case.

\subsection{Case 1}

Let the perturbation be carried out through
polytropic model parameters \begin{eqnarray}\label{40}
&&n\longrightarrow \tilde{n}+\delta n,
~~~q\longrightarrow \tilde{q}+\delta q,
~~~h\longrightarrow \tilde{h}+\delta h,
\end{eqnarray}
assuming that radial pressure remain same
after perturbation, then from Eq.$(\ref{14})$ we can write
\begin{equation}\label{41}
\tilde{P_r}=P_r=\alpha_1 \rho_{gc}\theta^n +
K \rho_{gc}^{1+\frac{1}{n}}\theta^{1+n}.
\end{equation}
Also from Eq.$(\ref{15})$
\begin{equation}\label{42}
\tilde{\rho}=\rho_{gc}\theta^{1+\tilde{n}}+\tilde{n}P_r.
\end{equation}
Thus the perturb form of Eq.$(\ref{30})$  become
\begin{equation}\label{43}
\hat{R_3}=\frac{d P_r}{dr}+\tilde{h}
(\tilde{\rho}+P_r)\Big(\frac{4\pi r^4 P_r-q^2+\tilde{m} r}
{r(r^2-2\tilde{m}r+q^2)}\Big)-\frac{\tilde{q}}
{4\pi r^4}\frac{d \tilde{q}}{dr}=0.
\end{equation}
Now using Eqs.$(\ref{16})$, $(\ref{41})$ and
$(\ref{42})$ in Eq.$(\ref{43})$, we get
\begin{eqnarray}\label{44}\notag
&&\tilde{R_3}=(n\alpha_1\theta^{n-1}+\alpha_3
 \theta^{n})\frac{d \theta}{d \xi}
- \frac{ \alpha_4 }{\alpha_3 \tilde{q}\xi^4}\frac{d \tilde{q}}{d \xi}
+\tilde{h}(\tilde{\theta^{n}}+(2\tilde{n}+1)\\&&
(\alpha_1\theta^{n}+\alpha\theta^{n+1}))\Big(\frac{(\alpha_1\theta^{n}+
\alpha\theta^{n+1})\xi^4 -\alpha_3^{-1}\alpha_4+\xi\tilde{v}(\xi)}
{\alpha_3^{-1}\xi^3-2\tilde{v}(\xi)\xi^2-\alpha_3^{-1}\alpha_4\xi}\Big).
\end{eqnarray}
From the above equation it follows up to first order, we may write
\begin{eqnarray}\label{45}
\delta\hat{R_3}=\hat{R_3}\Big(\xi,1+\delta n,
 h+\delta h v+\delta v, q +\delta q\Big),
\end{eqnarray}
\begin{eqnarray}\label{46}\notag
\delta {R_3}&=&\Big(\frac{\partial {\tilde{R_3}}}
{\partial \tilde{n}}\Big)
\mathrel{\mathop{\Big|_{\tilde{n}=n,~\tilde{v}=v}}_
{\mathrm{\tilde{h}=h,~\tilde{q}=q}}} \delta n
+ \Big(\frac{\partial {\tilde{R_3}}}{\partial \tilde{v}}\Big)
\mathrel{\mathop{\Big|_{\tilde{n}=n,~\tilde{v}=v}}_
{\mathrm{\tilde{h}=h,~\tilde{q}=q}}} \delta v\\
&&+ \Big(\frac{\partial {\tilde{R_3}}}
{\partial \tilde{h}}\Big) \mathrel{\mathop{\Big
|_{\tilde{n}=n,~\tilde{v}=v}}_{\mathrm{\tilde{h}=h,
~\tilde{q}=q}}} \delta h+ \Big(\frac{\partial {\tilde{R_3}}}
{\partial \tilde{q}}\Big) \mathrel{\mathop{\Big
|_{\tilde{n}=n,~\tilde{v}=v}}_{\mathrm{\tilde{h}=h,~\tilde{q}=q}}} \delta q.
\end{eqnarray}
Now suppose that $\beta=\{-2 v \xi ^2+\frac{\xi ^3}
{\alpha_3}+\frac{\alpha_4}{\alpha_3 }\xi\}^{-1}$ and using Eq. $(\ref{44})$, we obtained
\begin{eqnarray}\label{47}\notag
\frac{\partial {\tilde{R_3}}}{\partial \tilde{n}}
\mathrel{\mathop{\Big|_{\tilde{n}=n,~\tilde{v}=v}}_
{\mathrm{\tilde{h}=h,~\tilde{q}=q}}} &=&
\beta h(\theta ^n \text{Log}[\theta ]+2  (\alpha
 \theta ^{1+n}+\theta ^n \alpha _1 ) ) (v \xi -\frac{\alpha_4}
 {\alpha_3 }\\&& +\xi ^4  (\alpha  \theta ^{1+n}+\theta ^n \alpha _1 ) ) ,
\end{eqnarray}
\begin{eqnarray}\label{48}\notag
&&\frac{\partial {\tilde{R_3}}}{\partial
\tilde{v}}\mathrel{\mathop{\Big|_{\tilde{n}=n,~\tilde{v}=v}}_
{\mathrm{\tilde{h}=h,~\tilde{q}=q}}}=
\beta^{2} 2 h \xi ^2 (\theta ^n+(1+2 n)  (\alpha  \theta ^{1+n}+\theta ^n
\alpha _1 ) )  (v \xi -\frac{\alpha_4}{\alpha_3 }
\\&&+\xi ^4  (\alpha  \theta ^{1+n}+\theta ^n \alpha _1 ) )+
\beta h \xi (\theta ^n+(1+2 n)  (\alpha  \theta ^{1+n}+\theta ^n \alpha _1 ) )
\end{eqnarray}
\begin{eqnarray}\label{49}\notag
\frac{\partial {\tilde{R_3}}}{\partial \tilde{h}}
\mathrel{\mathop{\Big|_{\tilde{n}=n,~\tilde{v}=v}}_
{\mathrm{\tilde{h}=h,~\tilde{q}=q}}}
&=&\beta(\theta ^n+(1+2 n)  (\alpha
\theta ^{1+n}+\theta ^n \alpha _1 ) ) (v \xi -\frac{\alpha_4}{\alpha_3}
\\&&+\xi ^4 (\alpha  \theta ^{1+n}+\theta ^n \alpha _1 ) ),
\end{eqnarray}
\begin{eqnarray}\label{50}\notag
\frac{\partial {\tilde{R_3}}}{\partial \tilde{q}}
\mathrel{\mathop{\Big|_{\tilde{n}=n,~\tilde{v}=v}}_
{\mathrm{\tilde{h}=h,~\tilde{q}=q}}} &=&
-\frac{8 h \pi  q  P_{\text{rc}}}{\alpha_3^2 \alpha}\Big\{
\beta(\theta ^n+(1+2 n)  (\alpha  \theta ^{1+n}+\theta ^n \alpha _1 ) )+
\\ \notag &&
\beta^2 \xi (\theta ^n+(1+2 n)  (\alpha  \theta ^{1+n}+\theta ^n \alpha _1 ) )
(v \xi -\frac{\alpha_4}{ \alpha_3}+\\&&\xi ^4
 (\alpha  \theta ^{1+n}+\theta ^n \alpha _1 ) )
 \Big\}-\frac{4 \pi  P_{\text{rc}} \frac{d q}{d \xi}}{\alpha_3^2 \alpha \xi ^4}
-\frac{4 \pi  q P_{\text{rc}} \frac{d^2 q}{d \xi^2} }{\alpha_3^2 \alpha \xi ^4}.
\end{eqnarray}
Also From Eq.$(\ref{19})$, we have
\begin{eqnarray}\label{51}
 \tilde{v}=\int_0^{\xi} \Big[\bar{\xi}^2
 \big\{\theta^{\tilde{n}}+\tilde{n}
 (\alpha_1\theta^{n}+\alpha\theta^{n+1})\big\}
 -\frac{\alpha_4}{\alpha_3 \bar{\xi}^2}
 +\frac{\alpha_4}{\alpha_3 \bar{\xi} q}\frac{d q}{d \xi}\Big]d \bar{\xi},
\end{eqnarray}
and
\begin{eqnarray}\label{52}
\delta v=F_1 \delta n,~~~\delta
 q=\frac{F_1}{F_2}\delta n~~~\delta h=-\Gamma\delta n,
\end{eqnarray}
where
\begin{eqnarray}\label{53}\notag
&&F_1=\int_0^{\xi}  \bar{\xi}^2\big\{\theta^{\tilde{n}}log
 \theta + (\alpha_1\theta^{n}+\alpha\theta^{n+1})\big\}
  d \bar{\xi},
\\&&  F_2 =\int_0^{\xi} \frac{\alpha_4}{\alpha_3 \bar{\xi} q}
\Big[ -\frac{1}{q \bar{\xi}}+\frac{d q}{d \xi}+q\frac{d^2 q}{d \xi^2}\Big]d \bar{\xi},
\end{eqnarray}
and
\begin{eqnarray}\label{54}
\Gamma=\Bigg(
\frac{F_1\frac{\partial\hat{\tilde{R}}}{\partial \tilde{v}}
+F_2\frac{\partial\hat{\tilde{R}}}{\partial \hat{\tilde{q}}}
+\frac{\partial\hat{\tilde{R}}}{\partial \tilde{n}} }
{\frac{\partial\hat{\tilde{R}}}{\partial \tilde{h}}}\Bigg)
\mathrel{\mathop{\Big|_{\tilde{n}=n,~\tilde{v}=v}}
_{\mathrm{\tilde{h}=h,~\tilde{q}=q}}}.
\end{eqnarray}
So
\begin{eqnarray}\label{55}\notag
\delta R_3 &=&
 \Bigg( \beta h  (\theta ^n \text{Log}[\theta ]+2  (\alpha  \theta ^{1+n}+\theta ^n \alpha _1 ) )
 (v \xi -\frac{\alpha_4}{\alpha_3}+\xi ^4  (\alpha  \theta ^{1+n}+\theta ^n \alpha _1 ) )
\\ \notag &-&
 \Gamma \beta  (\theta ^n+(1+2 n) (\alpha
 \theta ^{1+n}+\theta ^n \alpha _1 ) )
 (v \xi -\frac{\alpha_4}{\alpha_3}+\xi ^4
 (\alpha  \theta ^{1+n}+\theta ^n \alpha _1 ) )
\\ \notag &+& F_1
\Big(
h\beta \xi   (\theta ^n+(1+2 n)  (\alpha
 \theta ^{1+n}+\theta ^n \alpha _1 ) )
+2h\beta^2 \xi^2 (\theta ^n+(1+2 n) \\ \notag &&
 (\alpha  \theta ^{1+n}+\theta ^n \alpha_1 ) )  (v \xi -
 \frac{\alpha_4}{\alpha_3}
 +\xi ^4 (\alpha  \theta ^{1+n}+\theta ^n \alpha _1 ) )
\Big)
\\ \notag &+&
\frac{F_1}{F_2}  \Big(
-\frac{8 h \pi  q P_{\text{rc}} }{\alpha_3^2\alpha}\Big(
 \beta (\theta ^n+(1+2 n)  (\alpha  \theta ^{1+n}
 +\theta ^n \alpha _1 ) )
\\ \notag &+& \xi \beta^2 (\theta ^n+(1+2 n)
(\alpha  \theta ^{1+n}+\theta ^n \alpha _1 ) )  (v \xi
-\frac{\alpha_4}{\alpha_3}+\xi ^4  (\alpha
  \theta ^{1+n}+\theta ^n \alpha _1 ) )\Big)
\\&&-
\frac{4 \pi  P_{\text{rc}} \frac{d q}{d \xi}}{\alpha
\alpha_3^2 \xi ^4}-\frac{4 \pi  q P_{\text{rc}}
\frac{d^2 q}{d \xi^2} }{\alpha \alpha_3^2 \xi ^4} \Big) \Bigg)\text{$\delta $n}.
\end{eqnarray}
It would be more convenient to use variable $x$ defined by
\begin{eqnarray}\label{56}
\xi= \bar{A}x,~~~\bar{A}=A r_\Sigma=\xi_\Sigma,
\end{eqnarray}
then
\begin{eqnarray}\label{57}\notag
\delta R_3 &=&
 \Bigg( \beta h  (\theta ^n \text{Log}[\theta ]+2
  (\alpha  \theta ^{1+n}+\theta ^n \alpha _1 ) )
 (v (\bar{A}x) -\frac{\alpha_4}{\alpha_3}+(\bar{A}x) ^4
   (\alpha  \theta ^{1+n}+\theta ^n \alpha _1 ) )
\\ \notag &-&
 \Gamma \beta  (\theta ^n+(1+2 n) (\alpha  \theta ^{1+n}
 +\theta ^n \alpha _1 ) )  (v (\bar{A}x) -\frac{\alpha_4}
 {\alpha_3}+(\bar{A}x) ^4  (\alpha  \theta ^{1+n}+\theta ^n \alpha _1 ) )
\\ \notag &+& F_1
\Big(
h\beta (\bar{A}x)   (\theta ^n+(1+2 n)
(\alpha  \theta ^{1+n}+\theta ^n \alpha _1 ) )
+2h\beta^2 (\bar{A}x)^2 (\theta ^n+(1+2 n) \\ \notag &&
 (\alpha  \theta ^{1+n}+\theta ^n \alpha_1 ) )  (v (\bar{A}x) -
 \frac{\alpha_4}{\alpha_3}
 +(\bar{A}x) ^4 (\alpha  \theta ^{1+n}+\theta ^n \alpha _1 ) )
\Big)
\\ \notag &+&
\frac{F_1}{F_2}  \Big(
-\frac{8 h \pi  q P_{\text{rc}} }{\alpha_3^2\alpha}\Big(
 \beta (\theta ^n+(1+2 n)  (\alpha  \theta ^{1+n}+\theta ^n \alpha _1 ) )
\\ \notag &+& (\bar{A}x) \beta^2 (\theta ^n+(1+2 n)
 (\alpha  \theta ^{1+n}+\theta ^n \alpha _1 ) )  (v (\bar{A}x)
-\frac{\alpha_4}{\alpha_3}+(\bar{A}x) ^4  (\alpha
\theta ^{1+n}+\theta ^n \alpha _1 ) )\Big)
\\&&-
\frac{4 \pi  P_{\text{rc}} \frac{d q}{d x}}{\alpha
\alpha_3^2 \bar{A}^5 x^4}-\frac{4 \pi  q P_{\text{rc}}
  \frac{d^2 q}{d x^2} }{\alpha \alpha_3^2 \bar{A}^6 x^4} \Big) \Bigg)\text{$\delta $n}.
\end{eqnarray}
We will use above equation to plot the perturbed force  $\frac{\delta R_3}{\delta n}$ against radius of
star and observe it for possible occurrence of cracking (overturning) in polytropes of first kind developed under the GPEoS.
\begin{figure} \label{fig7}
\centering
\includegraphics[width=80mm]{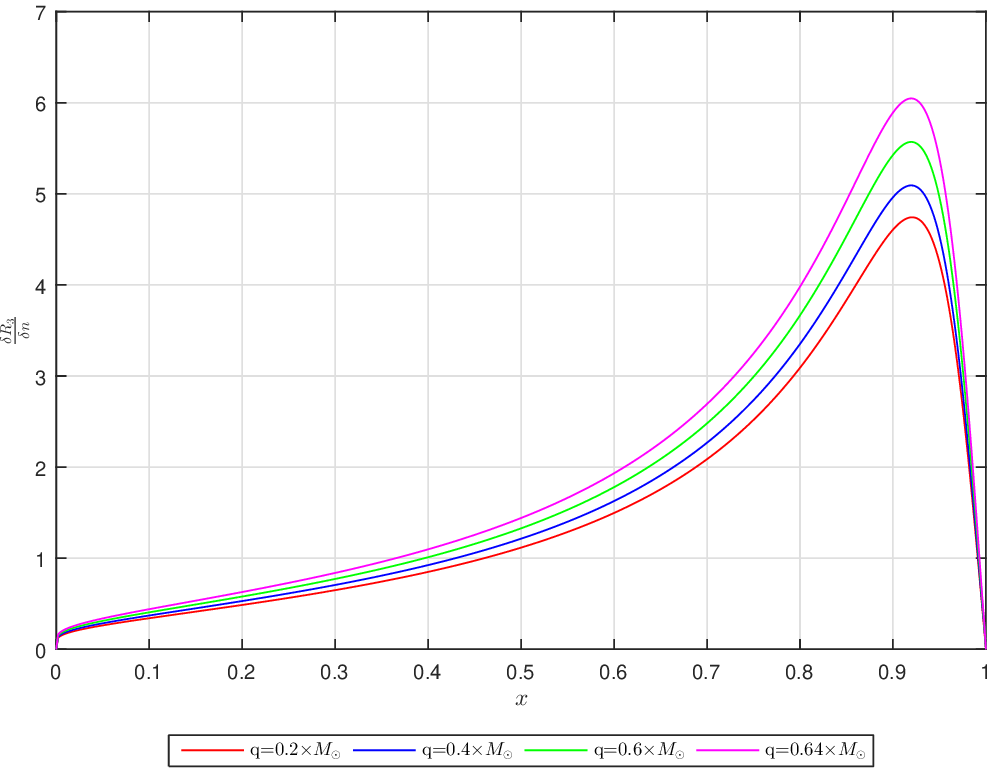}
\caption{Case $1$: Perturbation through n, q and h. $\frac{\delta R_3}{\delta n}$ as a function of $x$ for $n=1,~\alpha=8 \times 10 ^{-11},~\alpha_1=0.2, h=1.5, \Gamma=1.4$,
red curve: q=0.2 $M_\odot$,
blue curve: q=0.4 $M_\odot$,
green curve: q=0.6 $M_\odot$,
magenta curve: q=0.64 $M_\odot$}.
\end{figure}
\begin{figure} \label{fig8}
\centering
\includegraphics[width=80mm]{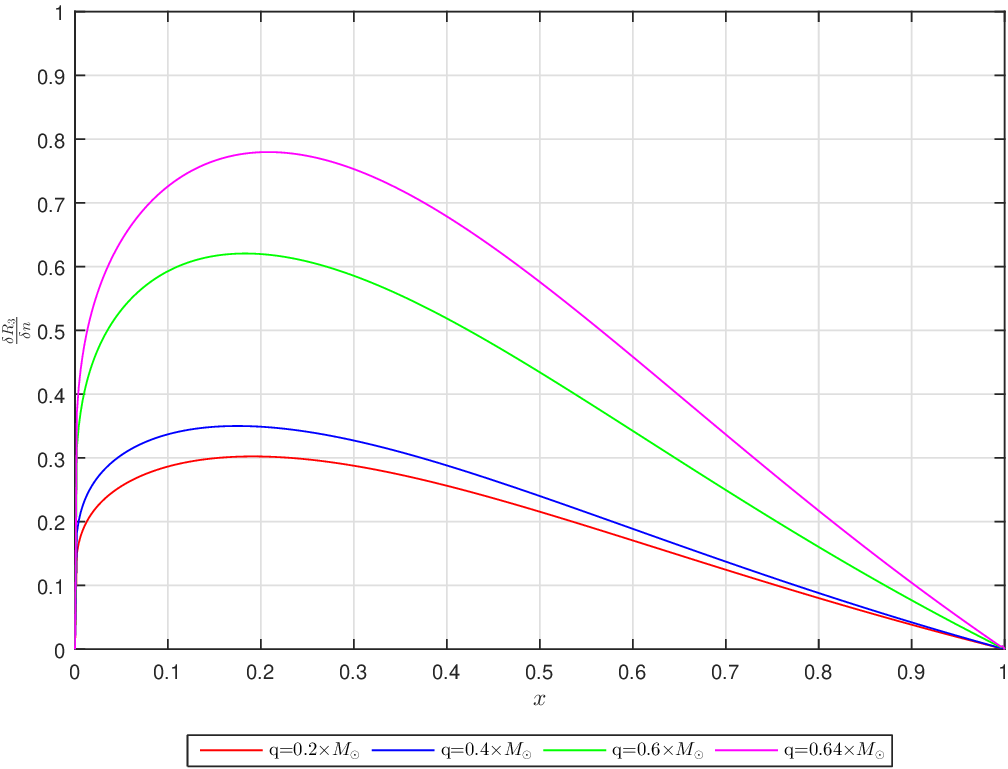}
\caption{Case $1$:  Perturbation through n, q and h. $\frac{\delta R_3}{\delta n}$ as a function of $x$ for $n=1,~\alpha=2 \times 10 ^{-10},~\alpha_1=0.4, h=0.5, \Gamma=1.0$,
red curve: q=0.2 $M_\odot$,
blue curve: q=0.4 $M_\odot$,
green curve: q=0.6 $M_\odot$,
magenta curve: q=0.64 $M_\odot$}.
\end{figure}
\begin{figure} \label{fig9}
\centering
\includegraphics[width=80mm]{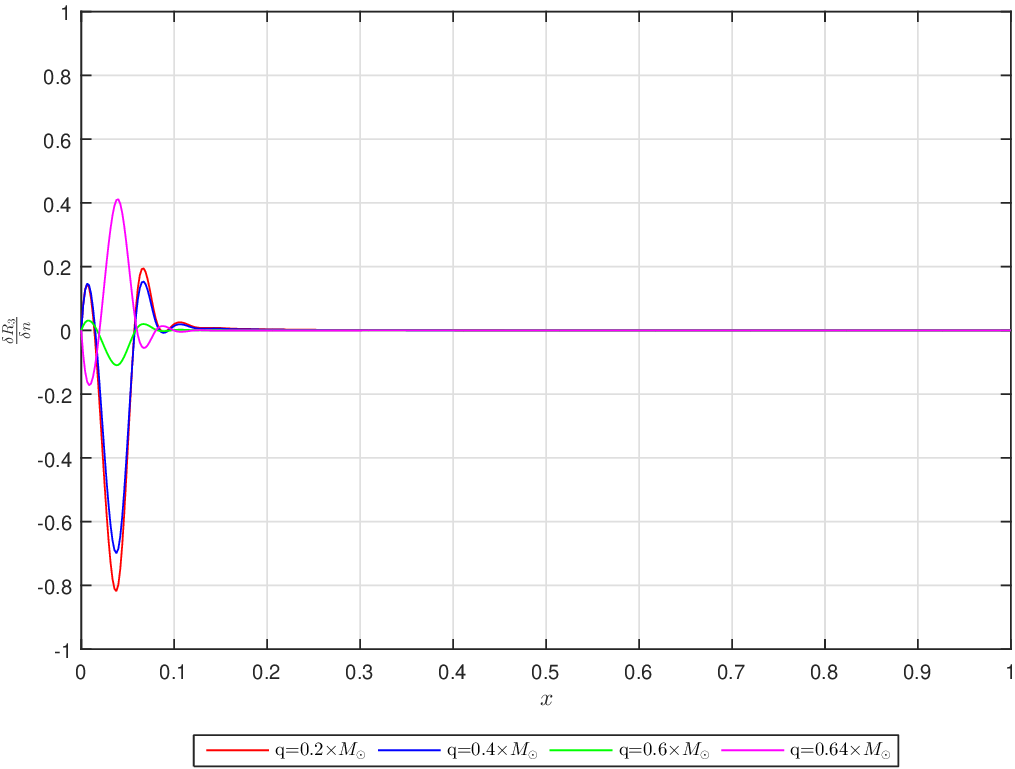}
\caption{Case $1$:  Perturbation through n, q and h. $\frac{\delta R_3}{\delta n}$ as a function of $x$ for $n=1,~\alpha=0.8,~\alpha_1=0.2, h=1.5, \Gamma=1.5$,
red curve: q=0.2 $M_\odot$,
blue curve: q=0.4 $M_\odot$,
green curve: q=0.6 $M_\odot$,
magenta curve: q=0.64 $M_\odot$}.
\end{figure}
\begin{figure} \label{fig10}
\centering
\includegraphics[width=80mm]{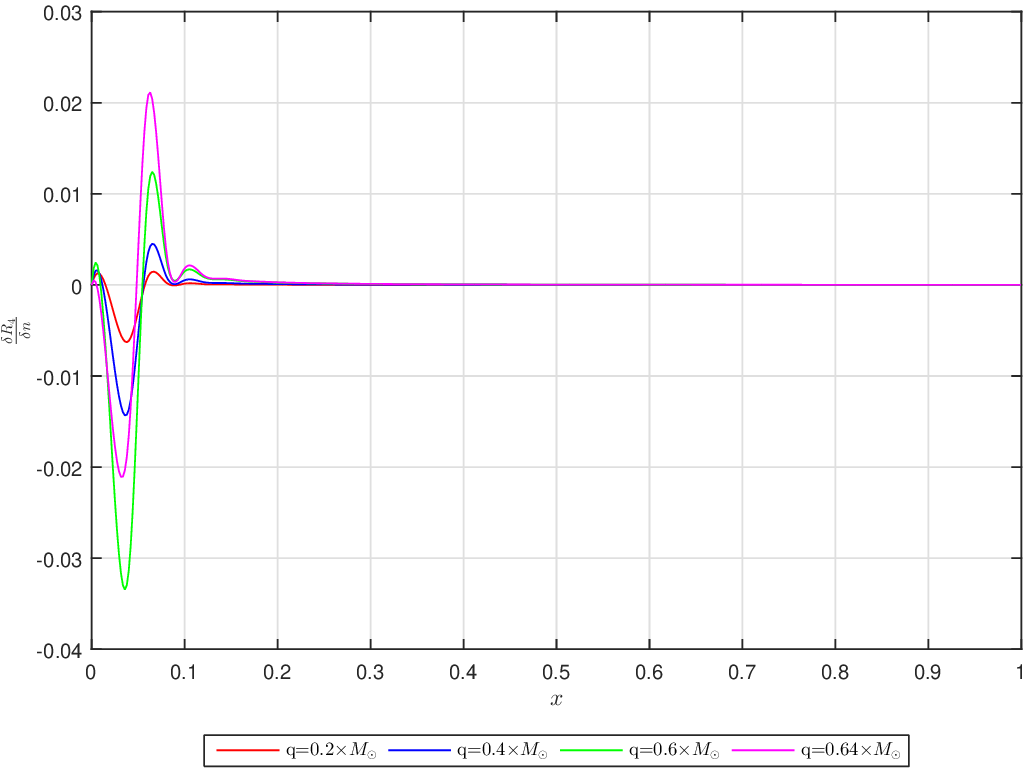}
\caption{Case $2$: Perturbation through n, q and h. $\frac{\delta R_4}{\delta n}$ as a function of $x$ for $n=1,~\alpha=8 \times 10 ^{-11},~\alpha_1=0.2, h=1.5, \Gamma=1.4$,
red curve: q=0.2 $M_\odot$,
blue curve: q=0.4 $M_\odot$,
green curve: q=0.6 $M_\odot$,
magenta curve: q=0.64 $M_\odot$}.
\end{figure}
\begin{figure} \label{fig11}
\centering
\includegraphics[width=80mm]{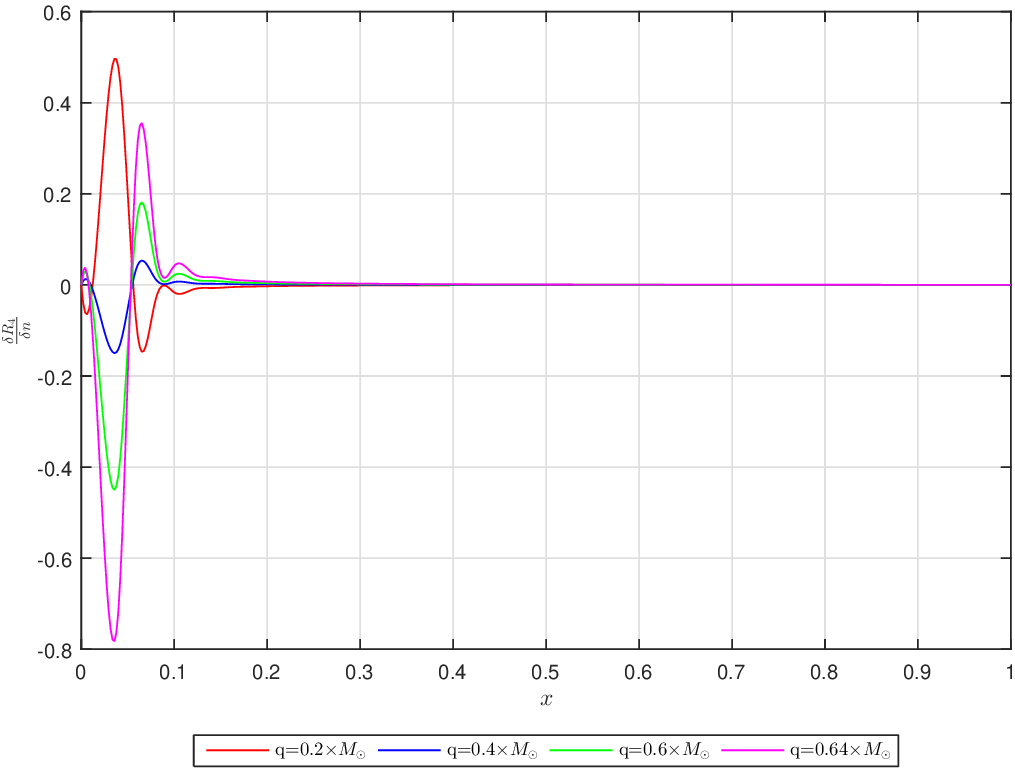}
\caption{Case $2$:  Perturbation through n, q and h. $\frac{\delta R_4}{\delta n}$ as a function of $x$ for $n=1,~\alpha=2 \times 10 ^{-10},~\alpha_1=0.4, h=0.5, \Gamma=1.0$,
red curve: q=0.2 $M_\odot$,
blue curve: q=0.4 $M_\odot$,
green curve: q=0.6 $M_\odot$,
magenta curve: q=0.64 $M_\odot$}.
\end{figure}

\subsection{Case 2}

Now we apply the parametric perturbation on polytropes of second kind here.
So from Eq.$(\ref{30})$ can be written as
\begin{eqnarray}\label{58}
\tilde{P_r}=P_r=\alpha_1 \rho_c \theta^n + K \rho_c^{1+\frac{1}{n}}\theta^{n+1}.
\end{eqnarray}
Now Eq.$(\ref{43})$ will transform as
\begin{eqnarray}\label{59}\notag
&&\tilde{R_4}=(n\alpha_1\theta^{n-1}+\alpha_3 \theta^{n})\frac{d \theta}{d \xi}
- \frac{ \alpha_4 }{\alpha_3 \tilde{q}\xi^4}\frac{d \tilde{q}}{d \xi}
+\tilde{h}(\tilde{\theta^{n}}+\alpha_1\theta^{n}+\alpha\theta^{n+1})
\\&&\Big(\frac{(\alpha_1\theta^{n}+
\alpha\theta^{n+1})\xi^4 -\alpha_3^{-1}\alpha_4+\xi\tilde{v}(\xi)}
{\alpha_3^{-1}\xi^3-2\tilde{v}(\xi)\xi^2-\alpha_3^{-1}\alpha_4\xi}\Big).
\end{eqnarray}
From the above equation it follows up to first order, we may write
\begin{eqnarray}\label{60}
\delta\hat{R_4}=\hat{R_4}\Big(\xi,1+\delta n,
 h+\delta h v+\delta v, q +\delta q\Big),
\end{eqnarray}
\begin{eqnarray}\label{61}\notag
\delta {R_4}&=&\Big(\frac{\partial {\tilde{R_4}}}
{\partial \tilde{n}}\Big)
\mathrel{\mathop{\Big|_{\tilde{n}=n,~\tilde{v}=v}}_
{\mathrm{\tilde{h}=h,~\tilde{q}=q}}} \delta n
+ \Big(\frac{\partial {\tilde{R_4}}}{\partial \tilde{v}}\Big)
\mathrel{\mathop{\Big|_{\tilde{n}=n,~\tilde{v}=v}}_
{\mathrm{\tilde{h}=h,~\tilde{q}=q}}} \delta v\\
&&+ \Big(\frac{\partial {\tilde{R_4}}}
{\partial \tilde{h}}\Big) \mathrel{\mathop{\Big
|_{\tilde{n}=n,~\tilde{v}=v}}_{\mathrm{\tilde{h}=h,
~\tilde{q}=q}}} \delta h+ \Big(\frac{\partial {\tilde{R_4}}}
{\partial \tilde{q}}\Big) \mathrel{\mathop{\Big
|_{\tilde{n}=n,~\tilde{v}=v}}_{\mathrm{\tilde{h}=h,~\tilde{q}=q}}} \delta q.
\end{eqnarray}
Then using Eq.$(\ref{59})$, we obtained
\begin{eqnarray}\label{62}
\frac{\partial {\tilde{R_4}}}{\partial \tilde{n}}
\mathrel{\mathop{\Big|_{\tilde{n}=n,~\tilde{v}=v}}_
{\mathrm{\tilde{h}=h,~\tilde{q}=q}}} =
\beta h \theta ^n \text{Log}[\theta ]
(v \xi -\frac{4 \pi  q^2 P_{\text{rc}}}{(1+n)^2
\alpha ^3}+\xi ^4  (\alpha  \theta ^{1+n}+\theta
^n \alpha _1 ) ),
\end{eqnarray}
\begin{eqnarray}\label{63}\notag
\frac{\partial {\tilde{R_4}}}{\partial \tilde{v}}
\mathrel{\mathop{\Big|_{\tilde{n}=n,~\tilde{v}=v}}_
{\mathrm{\tilde{h}=h,~\tilde{q}=q}}} &=&
\beta h \xi(\theta ^n+\alpha  \theta ^{1+n}+\theta ^n \alpha _1 )
+
2 \beta^2 h \xi ^2  (\theta ^n+\alpha  \theta ^{1+n}+\theta ^n \alpha _1 )
\\&& (v \xi -\frac{4
\pi  q^2 P_{\text{rc}}}{(1+n)^2 \alpha ^3}+
\xi ^4  (\alpha  \theta ^{1+n}+\theta ^n \alpha _1 ) ),
\end{eqnarray}
\begin{eqnarray}\label{64}
\frac{\partial {\tilde{R_4}}}{\partial \tilde{h}}
\mathrel{\mathop{\Big|_{\tilde{n}=n,~\tilde{v}=v}}_
{\mathrm{\tilde{h}=h,~\tilde{q}=q}}} =\beta
(\theta ^n+\alpha  \theta ^{1+n}+\theta ^n \alpha _1 )  (v \xi
-\frac{\alpha_4}{\alpha_3}+\xi^4  (\alpha  \theta ^{1+n}+\theta ^n \alpha _1 ) ),
\end{eqnarray}
\begin{eqnarray}\label{65}\notag
\frac{\partial {\tilde{R_4}}}{\partial \tilde{q}}
\mathrel{\mathop{\Big|_{\tilde{n}=n,~\tilde{v}=v}}_
{\mathrm{\tilde{h}=h,~\tilde{q}=q}}} &=&
\frac{8 h \pi  q  P_{\text{rc}}}{\alpha_3^2 \alpha}
\Big\{
\beta(\theta ^n+\alpha  \theta ^{1+n}+\theta ^n \alpha _1 )
-
\xi \beta^2(\theta ^n+\alpha  \theta ^{1+n}
\\ \notag &&+\theta ^n \alpha _1 )
(v \xi -\frac{4 \pi  q^2 P_{\text{rc}}}{(1+n)^2
\alpha ^3}+\xi ^4  (\alpha  \theta ^{1+n}+\theta ^n \alpha _1 ) )
\Big\}
\\&&
-\frac{4 \pi  P_{\text{rc}}
\frac{d q}{d \xi}}{(1+n)^2 \alpha ^3 \xi ^4}-\frac{4 \pi  q P_{\text{rc}}
\frac{d^2 q}{d \xi}^2}{(1+n)^2 \alpha ^3 \xi ^4}.
\end{eqnarray}
From Eq.$(\ref{33})$, we have
\begin{eqnarray}\label{66}
\tilde{v}=\int_0^{\xi}
\Big[\bar{\xi}^2 \theta^{\tilde{n}}
-\frac{\alpha_4}{\alpha_3
\bar{\xi}^2}+\frac{\alpha_4}
{\alpha_3 \bar{\xi} q}\frac{d q}{d \xi}\Big]d \bar{\xi},
\end{eqnarray}
and
\begin{eqnarray}\label{67}
\delta v=F_3 \delta n,~~~\delta
 q=\frac{F_3}{F_2}\delta n~~~\delta h=-\Gamma\delta n,
\end{eqnarray}
where
\begin{eqnarray}\label{68}
F_3=\int_0^{\xi}  \bar{\xi}^2 \theta^{\tilde{n}} log \theta d \bar{\xi},
\end{eqnarray}
and
\begin{eqnarray}\label{69}
\Gamma=\Bigg(
\frac{F_3\frac{\partial\hat{\tilde{R}}}{\partial \tilde{v}}
+F_2\frac{\partial\hat{\tilde{R}}}{\partial \hat{\tilde{q}}}
+\frac{\partial\hat{\tilde{R}}}{\partial \tilde{n}} }
{\frac{\partial\hat{\tilde{R}}}{\partial \tilde{h}}}\Bigg)
\mathrel{\mathop{\Big|_{\tilde{n}=n,~\tilde{v}=v}}
_{\mathrm{\tilde{h}=h,~\tilde{q}=q}}}.
\end{eqnarray}
So
\begin{eqnarray}\label{70}\notag
\delta R_4&=&\Bigg (\beta h \theta ^n
\text{Log}[\theta ]  (v \xi -\frac{\alpha_4}{\alpha_3}+\xi ^4 (\alpha  \theta ^{1+n}+\theta ^n \alpha _1 ) )
-\beta \Gamma   (\theta ^n+\alpha  \theta ^{1+n}
\\ \notag &&+\theta ^n \alpha _1 ) (v \xi -\frac{\alpha_4}{\alpha_3}+\xi ^4  (\alpha  \theta ^{1+n}+\theta ^n \alpha _1 ) )+
F_3  \Big( h \beta \xi   (\theta ^n+\alpha  \theta ^{1+n}
\\ \notag &&+\theta ^n \alpha _1 )+ 2\beta^2 h \xi ^2  (\theta ^n+\alpha  \theta ^{1+n}+\theta ^n \alpha _1 )
(v \xi -\frac{\alpha_4}{\alpha_3}+\xi ^4  (\alpha  \theta ^{1+n}
\\ \notag &&+\theta ^n \alpha _1 ) )  \Big)+\frac{F_3}{F_2}\Big(\frac{8 h \pi  q P_{\text{rc}}}{\alpha_3^2\alpha}\Big(\beta(\theta ^n+\alpha  \theta ^{1+n}+\theta ^n \alpha _1 )-\xi \beta^2 (\theta^n
\\ \notag &&+\alpha  \theta ^{1+n}+\theta ^n \alpha _1 )  (v \xi -\frac{4 \pi  q^2 P_{\text{rc}}}{(1+n)^2 \alpha ^3}+\xi ^4  (\alpha  \theta ^{1+n}+\theta
^n \alpha _1 ) )
\Big)
\\ &&-\frac{4 \pi  P_{\text{rc}} \frac{d q}{d \xi} }{(1+n)^2 \alpha ^3 \xi ^4}-\frac{4 \pi  q P_{\text{rc}} \frac{d^2 q}{d \xi^2} }{(1+n)^2 \alpha ^3 \xi ^4}
\Big)
\Bigg) \text{$\delta $n}.
\end{eqnarray}
It would be more convenient to use variable $x$ defined by
\begin{eqnarray}\label{71}
\xi= \bar{A}x,~~~\bar{A}=A r_\Sigma=\xi_\Sigma,
\end{eqnarray}
then
\begin{eqnarray}\label{72}\notag
\delta R_4&=&\Bigg (\beta h \theta ^n \text{Log}[\theta ]  (v (\bar{A}x) -\frac{\alpha_4}{\alpha_3}+(\bar{A}x) ^4 (\alpha  \theta ^{1+n}+\theta ^n \alpha _1 ) )
-\beta \Gamma   (\theta ^n+\alpha  \theta ^{1+n}
\\ \notag &&+\theta ^n \alpha _1 ) (v (\bar{A}x) -\frac{\alpha_4}{\alpha_3}+(\bar{A}x) ^4  (\alpha  \theta ^{1+n}+\theta ^n \alpha _1 ) )+
F_3  \Big( h \beta (\bar{A}x)   (\theta ^n+\alpha  \theta ^{1+n}
\\ \notag &&+\theta ^n \alpha _1 )+ 2\beta^2 h (\bar{A}x) ^2  (\theta ^n+\alpha  \theta ^{1+n}+\theta ^n \alpha _1 )
(v (\bar{A}x) -\frac{\alpha_4}{\alpha_3}+(\bar{A}x) ^4  (\alpha  \theta ^{1+n}
\\ \notag &&+\theta ^n \alpha _1 ) )  \Big)+\frac{F_3}{F_2}\Big(\frac{8 h \pi  q P_{\text{rc}}}{\alpha_3^2\alpha}\Big(\beta(\theta ^n+\alpha  \theta ^{1+n}+\theta ^n \alpha _1 )-(\bar{A}x) \beta^2 (\theta^n
\\ \notag &&+\alpha  \theta ^{1+n}+\theta ^n \alpha _1 )  (v (\bar{A}x) -\frac{4 \pi  q^2 P_{\text{rc}}}{(1+n)^2 \alpha ^3}+(\bar{A}x) ^4  (\alpha  \theta ^{1+n}+\theta
^n \alpha _1 ) )
\Big)
\\ &&-\frac{4 \pi  P_{\text{rc}} \frac{d q}{d  x} }{(1+n)^2 \alpha ^3 \bar{A}^5 x ^4}-\frac{4 \pi  q P_{\text{rc}} \frac{d^2 q}{d x^2} }{(1+n)^2 \alpha ^3 \bar{A}^6x ^4}
\Big)
\Bigg) \text{$\delta $n}.
\end{eqnarray}
We will use above equation to plot the perturbed force  $\frac{\delta R_4}{\delta n}$ against radius of
star and observe it for possible occurrence of cracking (overturning) in polytropes of first kind developed under the GPEoS.

\section{Conclusion and Discussion}

In this work, we have applied two different perturbation schemes on two types
of charged polytropes developed under the assumption of GPEoS \cite{31}.
In first perturbation scheme, we have used LDP scheme with conformally flat condition
and sketched the force distribution function against radius of star. Here, we have assumed that
all physical parameters involved in the model and their derivatives as
function of central density. Figure \textbf{1-3} show the plot of force distribution
function $\frac{\delta R_1}{\delta \rho_{gc}}$  against dimension less
radius $\xi$. It is observed that for different values of central
density and charge the system remain stable even after perturbation.
These plots represent a picture of system just after
perturbation. We observe rapid growth in the
perturbed force near the center but it become consistent, stable
and smooth as we move from center to outer surface of star.
The zoom box inside the figure depicted perturbed forces near the center.
It clearly indicated that the as the magnitude of charge
increases gradually the peak of perturbed forces become less comparatively
and it indicates that the presence charge stabilized the system after perturbation.
Such behavior is observed in both types of polytropes under LDP scheme (see figs. \textbf{1-6}).

In second perturbation scheme, we have perturbed the system through parameters
involved in the model like charge, anisotropy and polytropic index.
Such perturbation is carried out with the assumption that radial pressure
remain same even after perturbation. Under this perturbation scheme, the
perturbed force distribution $\frac{\delta R_3}{\delta n}$
is plotted against radius of star (see \textbf{7-9}). We get stable regions
for small values of $\alpha$ as shown in \textbf{7} and \textbf{8}. These plots also shows
that the system is very sensitive towards the choice of parameters. Initially the
perturbed force $\frac{\delta R_3}{\delta n}$ strong behavior near the boundary
of star (see Figure \textbf{7}) but slightly change in parameters shifts the magnitude of force
towards the center of star (see Figure \textbf{8}). In figs. \textbf{7} and \textbf{8} stable configurations
are observed. If the value of $\alpha$ is increased significantly then the system
become unstable after perturbation and cracking (overturning) is observed.
For smaller values of charge weak cracking near the center and strong overturning
occurs in the outer region. For sufficiently high value of charge we observe weak 
overturning near the center and strong cracking in the outer regions as shown in fig. \textbf{9}. 
Under parametric perturbation the second kind of polytropes did not show any
stable region and remain unstable under different combinations of parametric values.
Cracking and overturning is observed in polytropes of second type as shown in \textbf{10} and \textbf{11}.

From the above discussion, it is concluded that charged polytropes developed under 
GPEoS remain stable if LDP scheme is applied under conformally flat condition. 
They are not sensitive towards the perturbation in central density and such 
behavior appeared in both types discussed in this work. When perturbation is 
carried out through parameters, then first kind of polytropes show stable 
behavior for small values of $\alpha$ in the presence of charge but cracking 
(overturning) also appeared in this case for larger values. Further the second 
kind of polytropes remain unstable under under parametric perturbation. 
Hence the proper choice of parameters is very crucial in the study of polytropic models.

}
\end{document}